# Wireless Power Transfer Analysis using Ansys

Ziwei (Carl) Zhang
Supervisor: Xilin Liu
Department of Electrical and Computer Engineering, University of Toronto
ziweicarl.zhang@mail.utoronto.ca

*Abstract—* **In this report, methods of wireless power transfer of inside a cage is compared and simulated. Software Ansys Electronics Maxwell was used to simulate magnetic coupling of transmitter(s) and receiver with 10mA winding current in the mouse cage. Voltages of the device is not studied.**

I. Introduction

Wireless power transfer (WPT) is the process of which transmitter inductor coils transmit electrical power to a load across air group through electromagnetic field [1]–[3]. The energy is harvested by receiver coil, converting the electromagnetic field to energy before delivering to load attached to the transmitter. The load in the experimental setup is the electrode attached to a moving mouse in order to study mouse's brain activity [2], [4], [5].

In many modern applications, mobile phone wireless charging for example, require very close contact between transmitter and receiver; the slight misalignment causes charging process to stop or inefficient charging [6]–[8]. Wireless coverage is another key consideration, as the mouse cage system is designed to provide wireless charging in variable distance in the range of 10 cm in X and Y direction and 5 cm in Z direction [9], [10]. Due to the movement of animal, magnetic coupling between Tx and Rx coils should be modeled as variable with respect to distance between different Δx, Δy, Δz.

Despite that two-coil (Tx, Rx one each) inductive link yields optimized power transfer efficiency for matched impedance, it does not necessary deliver largest power to the load (PDL), while three-coil or four-coil inductive link provides extra degree of freedom in impedance matching, bring impedance load resistance closer to the targeted impedance for optimal power transfer efficiency [5], [11], [12]. As a result, this increases average coupling coefficient, which is also related to transmitter communication frequency. In the Ansys Electronics simulation 10 kHz is used as an example frequency, while Lee & Jia's paper mentioned 13.56 MHz as a possible frequency. Assuming a small source resistance of power source implemented in the design to provide maximum power to load, and that our cage powering uses low-carrier frequency, it is suggested in Lee & Jia's paper that two-coil or three-coil inductive link achieves greater power delivered to load and less variation of coupling coefficient. Henceforth, in the simulation study, only two-coil and three-coil are analyzed.

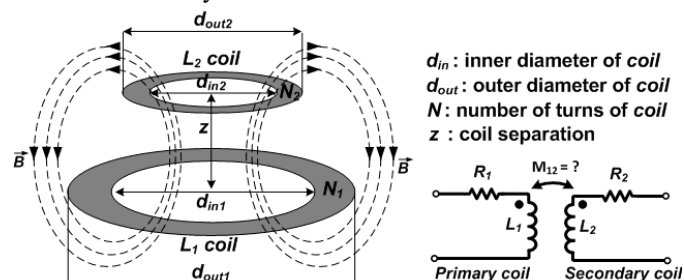

*Figure 1. Physical and electrical configurations in a two-coil inductive link. (Lee & Jia)*

Figure 1 shows physical configuration of two-coil, and future dimension specifications will be referenced using din_1, din_2, dout_1, dout_2, etc.

II. Physical and Electrical Setup

The entire system setup is simulated within a vacuum space which is hidden in active view. Solution type of eddy current is selected (Figure 2).

*A. Two-coil case study*

A transmitter coil and a receiver coil are used. To emulate the physical structure of transducer, aluminum and ferrite layers are added to the bare copper coils. The excitation condition and general coil parameters are summarized in Figure 3a and 3b (zoom size to view). Solid wires require only 1 wire, in the simulation stranded wires with 1000 windings are used for both transmitter and receiver coil. 10 mA eddy current with 90 degrees phase difference is applied to transmitter coil and receiver coil to simulate the maximum angular misalignment between the coils, thereby best theoretical power transfer; designer should consider worst scenario of angular misalignment in coupling coefficient estimated across the volume. One implementation suggested in the Lee & Jia's paper was omnidirectional powering that relies on resonators that direct magnetic flux towards receiver coil. Transmitter coil and receiver coil are parallel to each other in XY plane for max coupling coefficient, and 22 mm distance in Z axis; both coils are centered at x axis and self-symmetrical at X and Y axis, and are 0.5 mm thick. Transmitter coil has inner diameter (**din_1**) of 16 cm and outer diameter (**din_2**) of 30 cm; receiver coil has inner diameter

(**din_2**) of 8 mm and outer diameter (**dout_2**) of 20 mm. The dimension of coil size is provided by the coils used in the lab, and illustrated in the top view (Figure 4a) and side view (Figure 4b).

In the simulation, it is designed such that mouse have an active space of 6 cm in Z axis, and a total range of 16 cm in X or Y axis direction which means it moves X: [-8, 8] cm, Y: [-8, 8] cm. The symmetrical structure of system allows sweeping only in Z axis and one of X/Y axis, and is shown in Figure 5. However, a sweeping of all three variables can better estimate to completely cover minimsum coupling coefficients covered in the volume in contrast to a 2 D plot.

I obtained properties in rendering accuracies in mesh settings, mesh length size and elements restriction, solver convergence setting and passes through trial and errors, because of limitation in student version environment. By adjusting above parameters, results lose some accuracy compared to default conditions.

After simulation of parametric setup, I generated **EddyCurrent Report** of **3D Rectangular plots** from the results.

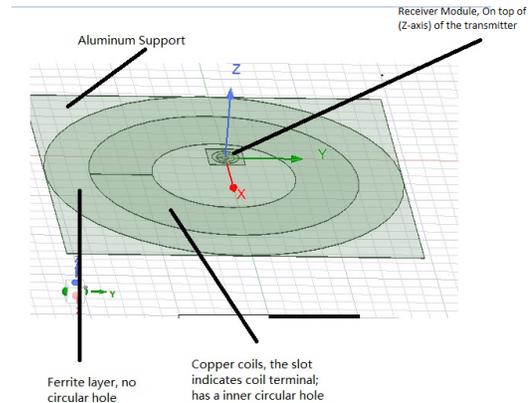

Figure 4a: Top view of 2-coil setup

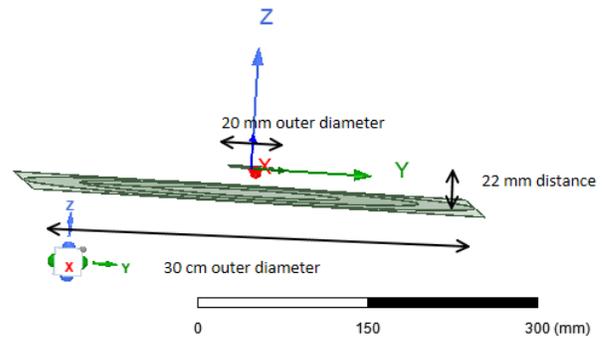

Figure 4b: Side view and dimension of two-coil setup

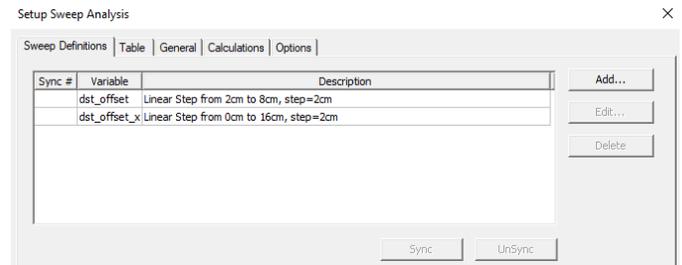

Figure 5: Sweeping parameters to emulate mouse movement. **dst_offset** represents movement in Z axis, **dst_offset_x** represents movement in X axis.

## B. Three-coil case study

Two transmitter coils and one receiver coil are used in three-coil simulation. The physical experimental setup is shown in Figure 6 provided by Professor Liu. In Ansys Electronics, the transmitter coils and receiver coil used the same dimensions as two-coil case, and the distance in Z-axis between transmitter and receiver coil remains 2 cm. Both transmitter coils are equal-sized, symmetrical along X-axis and parallel to XY plane; however, they are placed 6 mm (Z = -2 mm and Z = -8 mm) apart in Z-axis to prevent overlapping, and are centered at X = 12 cm and X = -12 cm respectively. The system is also simulated within a vacuum environment, but ferrite and aluminum layers of all three modules are removed to match the physical experimental setup which only include a copper layer. The simulation circuit is demonstrated in Figure 7, and excitation properties in Figure 8. A hundred instead of a thousand conductor windings were used for each terminal; The eddy currents are all 10 mA in magnitude, but the transmitter coil eddy currents are 60 degrees and 120 degrees leading receiver eddy current. I did not use 90 degrees and 180 degrees because applying current at a negative direction of the receiver cancels out the eddy current and coupling coefficient. Mouse have an active space of 6 cm in Z axis, and a total range of 16 cm in X axis direction, although simulation only covers positive X direction. The symmetrical distribution means positive X and negative X direction movement gives same results.

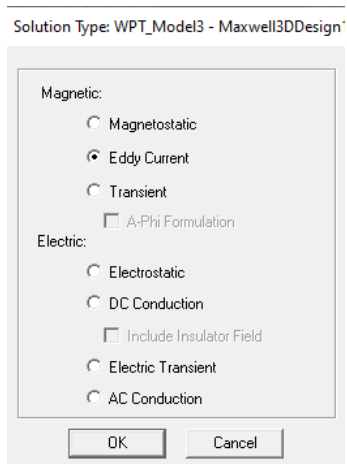

Fig 2: Solution Type of simulation environment.

Figure 3a: Two-coil terminal and Windings

Figure 3b: Two-coil Inductor Coil Parameters and Excitation Conditions

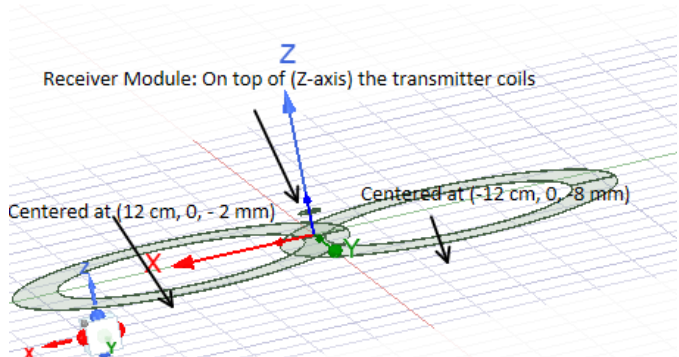

*Figure 7: Coil allocation of three-coil system*

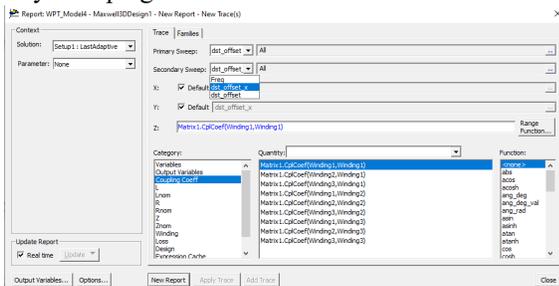

*Figure 8: Coil Terminals and windings in Excitations*

III. PLOTS OF COUPLING COEFFICIENTS

The 3-D Rectangular plots of EddyCurrent studies can be generated by sweeping offset variables

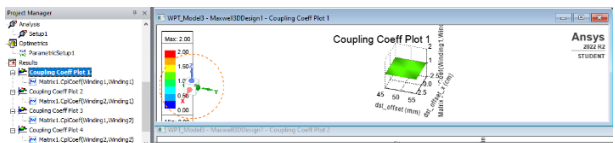

*Figure 9: Plot Coupling Coefficient between Windings*

A. Two-coil Simulation

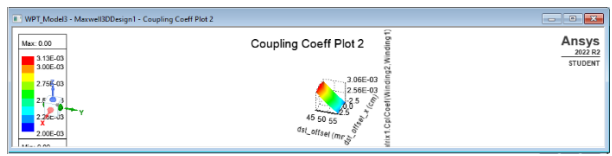

*Figure 10: Coupling coefficient to winding itself is always 1*

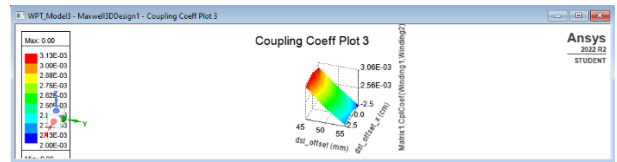

*Figure 11: Coupling coefficient between Winding 2 to 1 (X vs Z)*

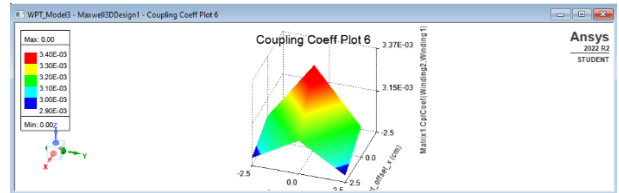

*Figure 12: Coupling coefficient between Winding 1 to 2 (X vs Z), shows that coupling coefficient between two windings is the same.*

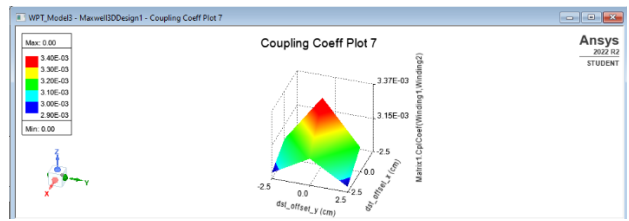

*Figure 13: Coupling coefficient between Winding 2 to 1 (X vs Y)*

*Figure 13: Coupling coefficient between Winding 1 to 2 (X vs Y), shows that coupling coefficient between two windings is the same.*

B. Three-Coil Simulation

After proving the symmetrical results of geometry in two-coil simulations, we only analyze offset X and Z positive direction sweeping, and plot one graph for coupling coefficient between two windings. Winding 1 matches with receiver terminal, while winding 2 and 3 match transmitter coil terminals.

Even with fewer conductor windings implemented, Three-coil simulation achieves higher coupling coefficient than two-coil. This is partially because of the location placement of winding terminal and the extra degree of freedom matching impedance. The two transmitter coils have highest coefficient of all because they are close to each other.

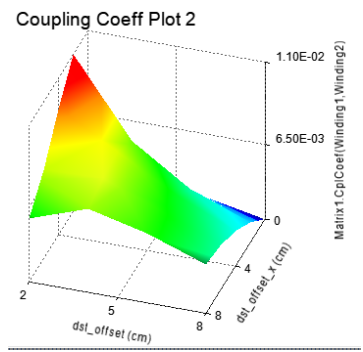

*Figure 14: Coupling coefficient between Winding 1 to 2*

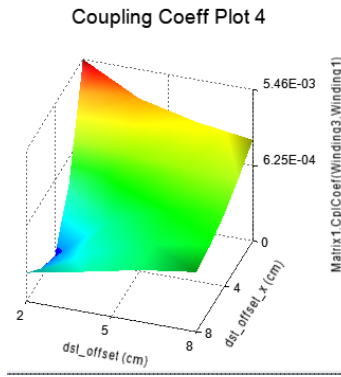

*Figure 15: Coupling coefficient between Winding 3 to 1*

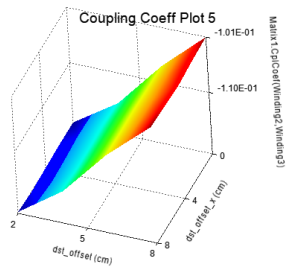

*Fig 16:* Coupling coefficient between Winding 2 to 3.

IV. CONCLUSION

Wireless Power Transfer can be simulated using Ansys Electronics software, by representing Tx, Rx coils with stranded wires (or solid wires) and applying eddy current of same magnitude but different phases. I followed tutorial Wireless Power Transfer Part I and Part II in constructing and modifying the models. It can help benchmark magnetic coupling of different geometries, materials of composition. However, simulation of power transfer, integrating microwave circuit model into **eddy-current** study exceeds the scope, and implementations should be validated experimentally. There is more information regarding testing in Burton et al paper, includes benchmark for voltage amplitudes and safety margins.